\documentclass[twocolumn]{aastex61}

\usepackage{graphicx}	% Including figure files
\usepackage{amsmath}	% Advanced maths commands
\usepackage{amssymb}	% Extra maths symbols

\newcommand{\Msun}{\,{\rm M_\odot}}
\newcommand{\Mblack}{M_\bullet}

\shorttitle{$M_{\bullet}-\sigma$ relation and luminosity function for low-mass black holes}
\shortauthors{Pacucci et al.}

\begin{document}

\title{Glimmering in the Dark: Modeling the low-mass end of the\\  $M_{\bullet}-\sigma$ relation and of the quasar luminosity function}

\correspondingauthor{Fabio Pacucci}
\email{fabio.pacucci@yale.edu}

\author[0000-0001-9879-7780]{Fabio Pacucci}
\affil{Yale University, Department of Physics, 
New Haven, CT 06511, USA}

\author{Abraham Loeb}
\affiliation{Harvard-Smithsonian Center for Astrophysics,
Cambridge, MA 02138, USA}

\author{Mar Mezcua}
\affiliation{Institute of Space Sciences (ICE, CSIC), Campus UAB, Carrer de Magrans s/n, E-08193 Barcelona, Spain}
\affiliation{Institut d'Estudis Espacials de Catalunya (IEEC), C/ Gran Capit\`a, E-08034 Barcelona, Spain}

\author{Ignacio Mart\'{i}n-Navarro}
\affiliation{University of California Santa Cruz, Santa Cruz, CA 95064, USA}
\affiliation{Max-Planck Institut f\"ur Astronomie, Konigstuhl 17, D-69117 Heidelberg, Germany}

\begin{abstract}
The $\Mblack - \sigma$ relation establishes a connection between central black holes (BHs) and their host spheroids. Supported by observations at $\Mblack \gtrsim 10^5 \Msun$, there is limited data on its validity at lower masses. Employing a semi-analytical model to simulate the combined evolution of BHs and their host galaxies, we predict the observational consequences of assuming a bimodality in the accretion efficiency of BHs, with low-mass BHs ($\Mblack \lesssim 10^5 \Msun$) accreting inefficiently. We predict a departure from the $\Mblack - \sigma$ relation at a transitional BH mass $\sim 10^5 \Msun$, with lower-mass BHs unable to reach the mass dictated by the relation and becoming disconnected from the evolution of the host galaxy. This prediction is an alternative to previous works suggesting a flattening of the relation at $\sim 10^5-10^6 \Msun$.  Furthermore, we predict a deficit of BHs shining at bolometric luminosities $\sim 10^{42} \, \mathrm{erg \, s^{-1}}$. Joined with a detection bias, this could partly explain the scarce number of intermediate-mass BHs detected. Conversely, we predict an increase in source density at lower bolometric luminosities, $<10^{42} \, \mathrm{erg \, s^{-1}}$. Because our predictions assume a bimodal population of high-redshift BH seeds, future observations of fainter BHs will be fundamental for constraining the nature of these seeds.

\end{abstract}


\section{Introduction} \label{sec:intro}
It is commonly accepted that the central region of all massive galaxies contains a super-massive black hole (BH, $\Mblack \gtrsim 10^6 \Msun$, see e.g. \citealt{King_Pounds_2015}). There seems to be a tight correlation between the mass of the BH and the properties of the host galaxy spheroid, such as the velocity dispersion of stars. This correlation, named the $\Mblack - \sigma$ relation \citep{Ferrarese_Merritt_2000, Gebhardt_2000, Kormendy_Ho_2013, McConnell_Ma_2013}, is surprising as there is a wide separation between the physical scale of the bulge of a galaxy and the sphere of influence of its central BH. The bulge of the Milky Way galaxy, for example, is $\sim 10^{4}$ times larger than the radius of influence of its BH.
The feedback resulting from BH accretion is thought to be the driving force in establishing the $\Mblack - \sigma$ relation, regulating both the star formation in massive host galaxies and the gas inflow onto the central BH \citep{Fabian_2000, Begelman_2005, King_Pounds_2015, Martin-Navarro_2018}.

\cite{Vdb_2016}, employing a heterogeneous set of 230 BHs with a minimum mass $\sim 4\times 10^5 \Msun$, found a relation of the form
\begin{equation}
\log \Mblack = (8.32\pm0.04) + (5.35\pm0.23)\log \sigma_{200} \, ,
\label{eq:M-sigma}
\end{equation}
where $\Mblack$ is in solar masses and $\sigma_{200}$ is expressed in units of $200 \, \mathrm{km \, s^{-1}}$.
Due to observational constraints, the low-mass regime of the relation is far less explored. Currently, the lightest central BH ($\Mblack \sim 3\times 10^4 \Msun$) is observed in a dwarf galaxy at $z \sim 0.03$ \citep{Chilingarian_2018}. Due to the paucity of the detected intermediate-mass BHs ($10^2 \Msun \lesssim  \Mblack \lesssim 10^6 \Msun$, e.g. \citealt{Greene_Ho_2004, Reines_2013, Baldassare_2015,Mezcua_2015,Mezcua_2016_X,Mezcua_2018_b}; see the review by \citealt{Mezcua_2017_review}), it is still hard to infer whether or not low-mass galaxies follow the extrapolation of the $\Mblack - \sigma$ relation \citep{Xiao_2011, Baldassare_2015, Mezcua_2017_review, MN_Mezcua_2018}.

A complete description of galaxy evolution requires a better understanding of the low-mass BH regime. Star formation and BH quenching in low-mass galaxies could be driven by different mechanisms, involving young stars and supernovae instead of the central BH \citep{Dubois_2015,Alcazar_2017,Habouzit_2017}.
For masses lighter than a transition mass, the central BH might be disentangled from the evolution of the host galaxy.

In this Letter we assume a bimodality in the accretion efficiency of BHs \citep{Pacucci_2017}, and predict the shape of the $\Mblack - \sigma$ relation and of the luminosity function for BHs with $\Mblack \lesssim 10^5 \Msun$. Our predictions, when compared to future observations of BHs in dwarf galaxies, will provide important constraints on the nature of BH seeds at high redshift, which constitute the progenitors of the $z \sim 7$ quasar population \citep{Fan_2006, Natarajan_2012, Volonteri_2016,Ricarte_2018}.

\section{A Bimodal Accretion Model} \label{sec:theory}

\cite{Pacucci_2017} suggested that accretion onto high-$z$ BHs may be bimodal. Accretion onto BHs lighter than a mass threshold $\widetilde{\Mblack}$ is inefficient, with largely sub-Eddington accretion rates and alternating quiescent and active phases.
Depending on the parameters of the model, $\widetilde{\Mblack} \sim 10^5-10^6 \Msun$. Previous studies already proposed that lower-mass BHs accrete more inefficiently than higher-mass ones (e.g., \citealt{PVF_2015, Pacucci_MaxMass_2017, Inayoshi_2016, Park_2016}). The novelty of the proposal by \cite{Pacucci_2017} was to identify the physical conditions that allow high-efficiency accretion. This identification allows to calculate the probability that a BH seed formed with an accelerated growth rate.

The high-efficiency region in the two-dimensional parameter space of BH mass and gas number density $(\Mblack, n_{\infty})$ is found by combining three conditions for efficient accretion on large $(r\gtrsim R_B)$ and small $(r\ll R_B)$ spatial scales, where $r$ is the distance from the BH and $R_B$ is its Bondi radius \citep{Bondi_1952}.
Assuming that photon trapping is active in the interior part of the accretion flow, the three conditions are as follows \citep{PVF_2015, Inayoshi_2016,Begelman_Volonteri_2017}.
The growth efficiency on small scales is
determined by the extent of the transition radius, above which the radiation pressure dominates the accretion flow: 
\begin{equation}
M_{\bullet} > 10^{-11}  \left(\frac{n_{\infty}}{1 \, \mathrm{cm^{-3}}}\right)^2 \, \mathrm{\Msun} \, .
\label{eq:small_scales_final}
\end{equation}
By increasing $n_{\infty}$ the minimum seed mass required to sustain efficient growth increases as well.
The growth efficiency on large scales is
determined by the comparison between $R_B$ and the extent of the ionized region around the BH:
\begin{equation}
M_{\bullet} > 10^9 \left( \frac{n_{\infty}}{\mathrm{1 \, cm^{-3}}} \right)^{-1} \, \mathrm{\Msun} \, .
\label{eq:large_scales_final}
\end{equation}
By increasing $n_{\infty}$ the minimum seed mass required to sustain efficient growth decreases.
Finally, the infalling gas needs to overcome the angular momentum barrier in order to accrete onto the BH. The condition is
\begin{equation}
M_{\bullet} > 2.2 \times 10^{19} \left( \frac{n_{\infty}}{\mathrm{1 \, cm^{-3}}} \right)^{-5/4} \lambda_{\rm B}^{3} \, \mathrm{\Msun} \, .
\label{eq:angmom_final}
\end{equation}
Here, $\lambda_{\rm B}$ is the ratio of the specific angular momentum of the gas $\ell_{\rm B}$ to its Keplerian value, computed at the trapping radius (distance from the BH inside which photon trapping is efficient): $\lambda_{\rm B} = \ell_{\rm B}/(G\Mblack R_{\rm B})^{1/2}$, where $G$ is the gravitational constant.

The minimum mass of a BH to be inside the high-efficiency region of the $(\Mblack, n_{\infty})$ parameter space is
\begin{equation}
\Mblack \gtrsim 5\times 10^5 \left( \frac{\lambda_{\rm B}}{10^{-1}} \right)^{24/13} \Msun \, .
\end{equation}
At higher BH masses, the BH is in the high-efficiency region for an increasingly larger range of gas density: high-efficiency accretion is, thus, more likely to occur.

\subsection{A mass threshold for super-Eddington rates}

\begin{figure}
\includegraphics[angle=0,width=0.5\textwidth]{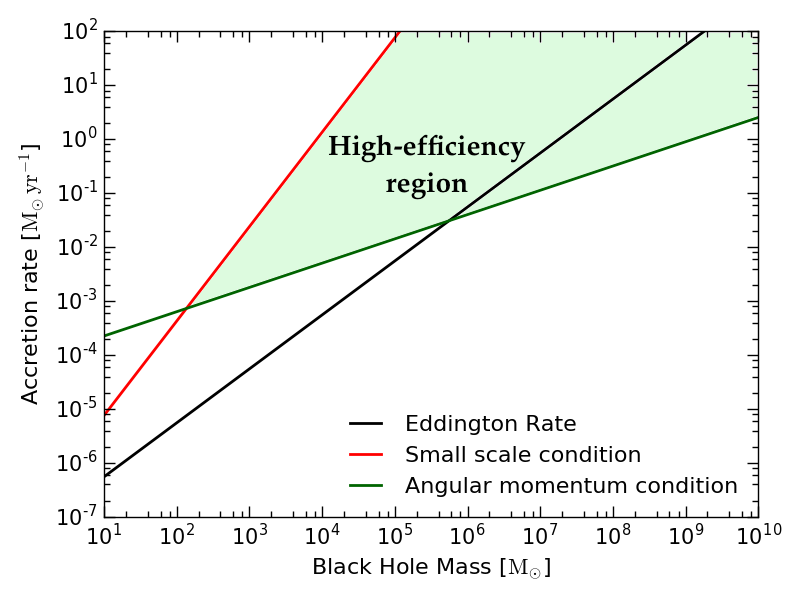}
\vspace{-0.4cm}
\caption{Conditions for efficient BH growth in the $(\Mblack , \dot{M}_B)$ parameter space. The high-efficiency region is shaded green; the Eddington rate is shown as a black line.}
\label{fig:theory}
\end{figure}

We present here a simplified argument to show that a fundamental transition between low-efficiency and high-efficiency accretion occurs in the  BH mass range $10^5 - 10^6 \Msun$, an assumption that is at the core of our bimodal model. The simplified assumptions introduced in this section are in no way used in the actual growth model described in Sec. \ref{sec:methods}.
We eliminate the parameter $n_{\infty}$ from Eqs. (\ref{eq:small_scales_final}) and (\ref{eq:angmom_final}) by computing the corresponding Bondi rate $\dot{M}_B = 4\pi \rho G^2 \Mblack^2/c_s^3$, 
where $\rho$ is the gas mass density, and $c_s$ is the sound speed. The Bondi rate is a convenient approximation to use in this simplified model; at scales $\gtrsim R_B$ is also a reasonable one for the accretion rate, as the accretion disk forms at much smaller scales. Assuming $c_s = \sigma_g \sim \sigma_s$ ($\sigma_g$ and $\sigma_s$ are the velocity dispersions for gas and stars, respectively; this simplified model remains valid as long as $\sigma_g$ and $\sigma_s$ are within the same order of magnitude) and that the $\Mblack-\sigma$ relation (Eq. \ref{eq:M-sigma}) is valid, the constraints of Eqs. (\ref{eq:small_scales_final}) and (\ref{eq:angmom_final}) in the $(\Mblack, \dot{M}_B)$ parameter space are shown in Fig. \ref{fig:theory}.

Beginning the growth at $\Mblack \ll 10^5 \Msun$, a BH seed can reach, at most, the Eddington rate. Growing in mass, it gets progressively closer to the high-efficiency regime, being able to enter it when the Eddington rate crosses the angular momentum barrier shown in Fig. \ref{fig:theory} as a green line. Once inside the high-efficiency region, super-Eddington rates are reachable \citep{Begelman_Volonteri_2017}.
The only condition that matters in this regime involves the angular momentum (Eq. \ref{eq:angmom_final}), whose relevance is not restricted to the high-$z$ Universe. An important assumption in this derivation is that there is always a gas supply to grow the BH at or above the Eddington limit: the growth is, thus, \textit{supply-limited}.

\section{Data and Methods} 
\label{sec:methods} 
Next, we predict the cosmological evolution of a population of BHs accreting in a bimodal regime. We compare our predictions with a sample of $\sim 300$ galaxies in the range $30 \lesssim \sigma(\mathrm{km \, s^{-1}}) \lesssim 400$.

\subsection{Description of the data}
\label{subsec:observations}
We test our model against observational data of both low-mass and high-mass galaxies for which BH mass and stellar velocity measurements are available. We use the same sample as that of \cite{MN_Mezcua_2018} (blue stars in Fig.~\ref{fig:M-sigma}), which includes a compilation of 127 low-mass Seyfert 1 galaxies with $\sigma<100$ km s$^{-1}$ from \cite{Xiao_2011} and \cite{Woo_2015} drawn from the Sloan Digital Sky Survey (SDSS; \citealt{SDSS_2018}). The BH masses of these galaxies were estimated from the width of optical broad emission lines under the assumption that the gas is virialized \citep{Xiao_2011,Woo_2015}. In the regime of massive galaxies, the sample includes 205 early-type galaxies with direct BH mass measurements from \cite{Vdb_2016} and $\sigma \gtrsim 100$ km s$^{-1}$. 
Typical errors in $\Mblack$ and $\sigma$ are up to a factor $\sim 3$ and $10-15 \, \mathrm{km \, s^{-1}}$, respectively.

\subsection{Modeling the $\Mblack-\sigma$ assuming bimodality}

We employ a merger tree code \citep{Parkinson_2008, Dayal_2017} to track the cosmological evolution of a population of dark matter halos distributed following the Sheth-Tormen halo mass function \citep{ST_1999}. We sample logarithmically the cosmological scale factor $a = (z+1)^{-1}$ between $z=20$ and $z=0.1$, the mean redshift of the sample described in Sec. \ref{subsec:observations}.
We assign a single BH seed to each $z=20$ galaxy with a halo mass $M_h \gtrsim 5 \times 10^7 \Msun$ (such that its virial temperature is higher than the atomic cooling threshold, see e.g. \citealt{BL01}). We further assume a ratio of high-mass ($\Mblack > 10^4 \Msun$) to low-mass ($\Mblack < 10^2 \Msun$) BH seeds of $1:100$. In fact, the formation of a high-mass seed is a much rarer event, because of the additional requirements (see e.g. \citealt{Bromm_Loeb_2003}) to prevent the fragmentation of the gas cloud. The ratio employed here is a proxy for the relative abundance of sources in the high-luminosity and the low-luminosity ends of the quasar luminosity function (see e.g. the one for $3<z<5$ in \citealt{Masters_2012}). This mixture of light and heavy seeds reproduces the $z \sim 0$ quasar luminosity function for $L_{\rm bol} \gtrsim 10^{43} \, \mathrm{erg \, s^{-1}}$ \citep{Hopkins_2007}, as detailed in Sec. \ref{subsec:LF}. We model high-mass seeds as direct collapse black holes (DCBHs; e.g., \citealt{Bromm_Loeb_2003}) and low-mass seeds as Pop III stellar remnants (e.g., \citealt{Hirano_2014}).
We model the initial mass function of DCBHs with a log-gaussian distribution, having a mean $\mu = 5.1$ and a standard deviation $\sigma = 0.2$, both in logarithm of mass.
For Pop III stars we employ a model with a Salpeter-like exponent and a low-mass cutoff $M_{c}$:
\begin{equation}
\Phi(\mathrm{Pop III},m) \propto m^{-2.35} \exp \left({-\frac{M_{c}}{m}} \right) \, .
\end{equation}
We assume $M_{c} = 10\, \mathrm{\Msun}$ and convolve this progenitor mass function with the relation \citep{Woosley_2002} between the mass of the remnant and the stellar mass. As long as there is a clear separation between the initial mass functions for low-mass and high-mass seeds, their exact shape does not significantly influence our results.

We calculate the central stellar velocity dispersion of the host from the asymptotic circular velocity $v_c$ (as a function of total halo mass and redshift), which is a proxy for the total mass of the dark matter halo of the galaxy: $\sigma = v_c/\sqrt{2}$.

The fueling of the central BH seed in each galaxy is implemented with the following scheme. A BH is \textit{active} whenever gas is available.
In the early Universe before a redshift threshold $z>z_t$, we assume that a sufficient amount of gas is always available to feed the seed: thus, the BH is always active. Simulations (e.g., \citealt{Dubois_2014}) and analytical estimates (e.g., \citealt{WL_2012}) suggest that even super-Eddington infall rates are fairly common at high redshift. \cite{Begelman_Volonteri_2017} point out that the fraction of AGNs accreting at super-Eddington rates could be as high as $\sim 10^{-3}$ at $z=1$ and $\sim 10^{-2}$ at $z=2$.
For $z<z_t$, we instead assume that the central BH is active only when a major merger occurs, defined as a merger with a mass ratio equal or larger than $1:10$. This criterion is meant to reflect the necessity of an external reservoir of gas to overcome the angular momentum barrier. Whenever a major merger occurs, the BH is set in the active phase for a time equal to the merger time scale \citep{Boylan-Kolchin_2008}.
We set $z_t = 6$ and check that our results remain qualitatively unchanged for all values of $z_t > 3$. We note, however, that the assumption of constant availability of gas for the central BHs is not realistic at these low $z$.

Whenever a BH is active, we describe the time evolution of $\Mblack$ with two parameters: the duty cycle ${\cal D}$ (fraction of the active phase spent accreting) and the Eddington ratio $f_{\rm Edd} = \dot{M}/\dot{M}_{\rm Edd}$. The first parameter describes the continuity of the gas inflow, and the second one quantifies the amount of mass flowing in.
The time evolution of $M_{\bullet}$, starting from $\Mblack(z_0)= M_{{\bullet},0}$, is obtained from the integral
\begin{equation}
M_{\bullet}(z) = M_{{\bullet},0} \exp \left(  \int_{z}^{z_0} \frac{{\cal C}({\cal D},f_{\rm Edd}) d{\rm z}}{{\cal E}(z)}  \right) \, ,
\label{eq:cosmo_evolution}
\end{equation}
where ${\cal C}({\cal D},f_{\rm Edd})$ incorporates various constants and the two parameters of the model, ${{\cal E}(z)}$ depends on the cosmology of choice (see \citealt{Pacucci_2017} for details).
The values adopted for ${\cal D}$ and $f_{\rm Edd}$ at each redshift depend on the properties of the BH in the $(\Mblack, n_{\infty})$ parameter space. In general, ${\cal D} \ll 1$ and $f_{\rm Edd} \ll 1$ in the low-efficiency region and ${\cal D} \sim 1$ and $f_{\rm Edd} \gtrsim 1$ in the high-efficiency region \citep{PVF_2015,Inayoshi_2016,Begelman_Volonteri_2017}. The gas density profile of the host galaxies is assumed to follow an isothermal sphere.

Once two galaxies merge, their central BHs are assumed to coalesce instantaneously. The only cap imposed for the growth by BH mergers is the mass of the most massive BHs thus far observed ($\sim 5\times 10^{10} \Msun$, \citealt{Mezcua_2018_a}). Expressed as a function of the velocity dispersion, the cap for the growth by accretion is devised to match the $\Mblack-\sigma$ relation for $\sigma \gtrsim 100 \, \mathrm{km \, s^{-1}}$. This mass cap is formally equal to Eq. (\ref{eq:M-sigma}). The relation $\Mblack \propto \sigma^{5.35}$ substantially agrees with the hypothesis of BH growth being regulated by energy-driven wind feedback (e.g., \citealt{King_2010_b}): $\Mblack \simeq f_g \kappa \sigma^5 /(\pi G^2 c)$, 
where $f_g$ is the cosmic baryon fraction and $\kappa$ is the gas opacity.

\section{OBSERVATIONAL PREDICTIONS } 
\label{sec:M-sigma}
We are now in a position to make observational predictions from our model, regarding the $\Mblack-\sigma$ relation and the quasar luminosity function.

\subsection{The low-mass regime in the $\Mblack-\sigma$}

\begin{figure*}
\centering
\includegraphics[angle=0,width=0.49\textwidth]{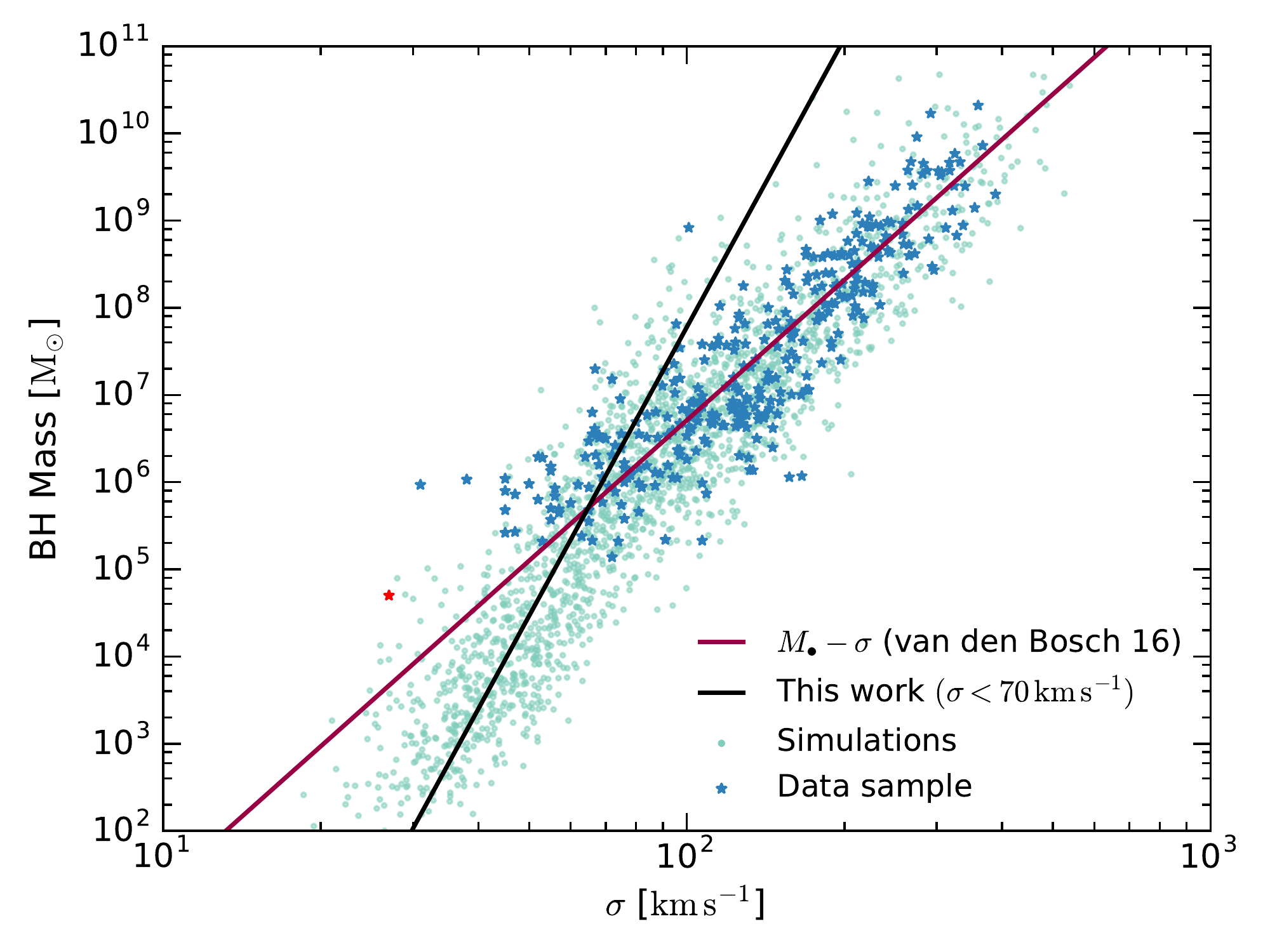}
\includegraphics[angle=0,width=0.49\textwidth]{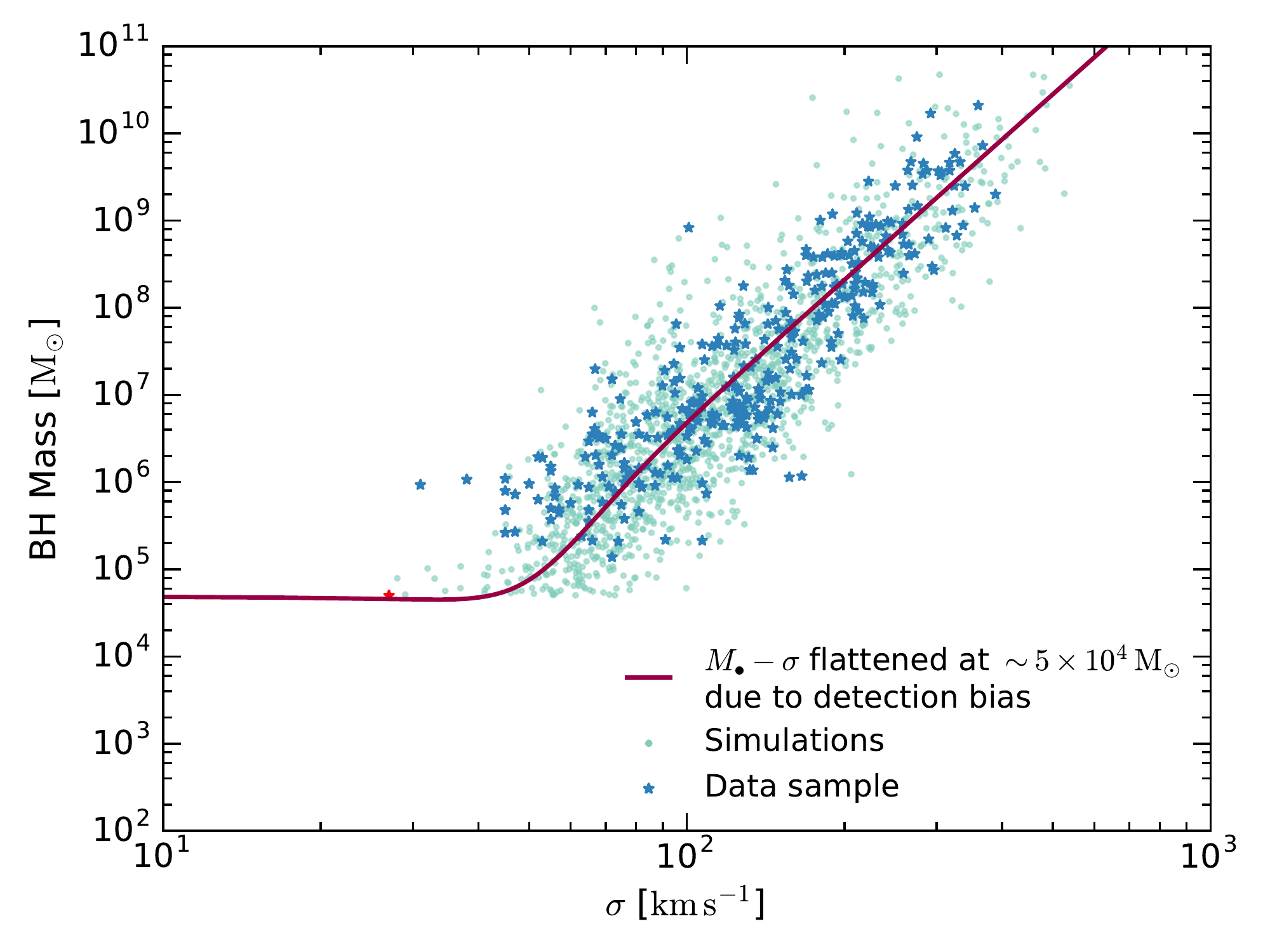}
\vspace{-0.4cm}
\caption{Theoretical predictions of the model (green points) compared with data from \cite{MN_Mezcua_2018} (blue stars). \textit{Left panel}: for $\Mblack \gtrsim 10^5 \Msun$ the simulations follow closely the well-known $\Mblack-\sigma$ relation (red line). For $\Mblack \lesssim 10^5 \Msun$ the simulations fall below the line, indicating that central BHs become disconnected from the evolution of their host stellar components. The \cite{Vdb_2016} model intercepts our predictions for $\Mblack \lesssim 10^5 \Msun$ at transition values $\sigma_t \sim 65 \, \mathrm{km \, s^{-1}}$ and $M_{\bullet,t} \sim 5 \times 10^5 \Msun$. \textit{Right panel}: current observational capabilities do not allow the detection of central BHs with $\Mblack \ll 10^5 \Msun$. For this reason we might be observing a flattening of the $\Mblack-\sigma$ relation toward masses $\Mblack \sim 10^5 - 10^6 \Msun$. The red line indicates a smooth transition in mass between the $\Mblack-\sigma$ relation at large BH masses and the observational limit at which both the BH mass and the stellar velocity dispersion are available (shown as a red star), currently set at $\Mblack \sim 5\times 10^4 \Msun$ \citep{Baldassare_2015}. The observational limit is interpreted here as a line of constant BH mass.}
\label{fig:M-sigma}
\end{figure*}

Our main results are shown in the left panel of Fig. \ref{fig:M-sigma}. The simulations are evolved to $z=0.1$ to match the mean redshift of the data sample in \cite{MN_Mezcua_2018}. 
The simulations closely follow the data and the theoretical model by \cite{Vdb_2016} for $\sigma > 70 \, \mathrm{km \, s^{-1}}$ (see also \citealt{MN_Mezcua_2018, Ricarte_2018}).
At $\sigma \sim 70 \, \mathrm{km \, s^{-1}}$ there is a clear change in trend: our simulation points tend to be under the theoretical $\Mblack-\sigma$ relation. This is a direct consequence of the bimodality in the accretion efficiency of BHs. For masses $\Mblack \gtrsim 10^5 \Msun$ the BHs are likely to be in the high-efficiency regime (see Fig. \ref{fig:theory}). These BHs grow efficiently and are able to reach the $\Mblack-\sigma$ within a Hubble time. When they reach the $\Mblack-\sigma$, their growth is saturated by energy-driven winds, which deplete the central regions from the remaining gas, preventing further growth. At this stage, growth can occur by mergers only. We find the presence of objects with masses significantly larger than the $\Mblack-\sigma$. Observationally, these objects can be interpreted as the brightest cluster galaxies, whose BHs are found to be over-massive with respect to the scaling relations \citep{Mezcua_2018_a}. Nonetheless, these objects constitute a minority population when compared to the bulk of BHs. 
For $\Mblack \lesssim 10^5 \Msun$, the fraction of BHs growing efficiently is small, and only a minority ($\sim 3\%$) of them are able to reach the cap imposed by the $\Mblack-\sigma$ relation. The vast majority of them remain at lower masses, unable to trigger sufficiently strong winds to fully halt their growth.
Instead, they keep accreting at very low rates for most of the Hubble time.
In the low-mass regime, we predict a much steeper relation ($\Mblack \sim \sigma^{11}$) which is only approximate, as BH and stellar component become increasingly disconnected.
The \cite{Vdb_2016} model intercepts our predictions for low masses at transition values $\sigma_t \sim 65 \, \mathrm{km \, s^{-1}}$ and $M_{\bullet,t} \sim 5 \times 10^5 \Msun$.
At lower masses and velocity dispersions \textit{the model clearly predicts a departure from the $\Mblack-\sigma$ relation, with the bulk of objects being under-massive}.

\subsection{The quasar luminosity function}
\label{subsec:LF}
\begin{figure}
\includegraphics[angle=0,width=0.5\textwidth]{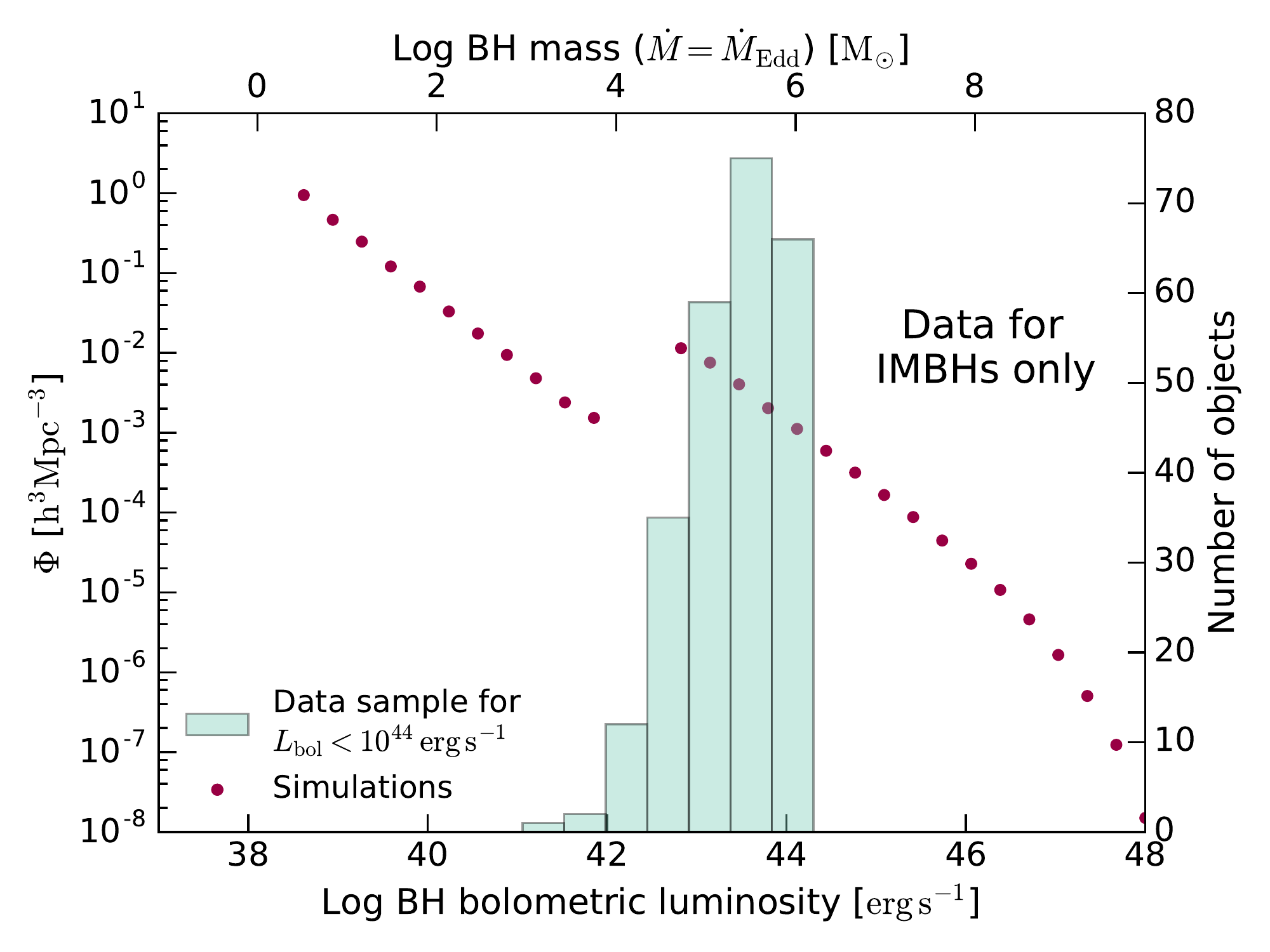}
\vspace{-0.4cm}
\caption{Luminosity function of galactic BHs in our simulations (red symbols), obtained from sources categorized in 35 logarithmic mass bins. The upper scale in BH mass assumes that all BHs are accreting at the Eddington rate and it is for reference only. There is a clear deficit of sources with predicted luminosities $L_{\rm bol} \sim 10^{42} \, \mathrm{erg \, s^{-1}}$. The data sample of intermediate-mass BHs from \cite{Greene_Ho_2004, Reines_2013,Baldassare_2017} also suggests this deficit, which might be a combination of observational bias and intrinsic scarcity of sources.}
\label{fig:LF}
\end{figure}

The simulated data at $z = 0.1$ can be used to construct a luminosity function, with the abundance of host halos retrieved from  a Sheth-Tormen halo mass function at the same redshift. We divide the simulations in $35$ logarithmic mass bins, and for each of them we compute the average luminosity. We then compare in Fig. \ref{fig:LF} our predictions with bolometric luminosities of BHs in dwarf galaxies, based on H$\alpha$ luminosities \citep{Greene_Ho_2004, Reines_2013,Baldassare_2017}. The bolometric correction is from \cite{Greene_Ho_2004}, $L_{\rm bol} = 2.34\times 10^{44}(L_{H\alpha}/10^{42})^{0.86} \, \mathrm{erg \, s^{-1}}$.
The luminosity function is in excellent agreement with observations for $L_{\rm bol} \gtrsim 10^{43} \, \mathrm{erg \, s^{-1}}$ (e.g., \citealt{Hopkins_2007}). At low luminosities there is an evident deficit around $L_{\rm bol} \sim 10^{42} \, \mathrm{erg \, s^{-1}}$, explained by our result that BHs accreting in that luminosity range have masses close to the transition mass between high and low efficiency. At higher BH masses, the likelihood of accreting close to the Eddington limit is large, while at lower masses the BH accretion is mostly sub-Eddington.
The result is a paucity of BHs shining at $L_{\rm bol} \sim 10^{42} \, \mathrm{erg \, s^{-1}}$. 
\textit{The scarcity of observations of intermediate-mass BHs in this luminosity range could thus be a combination of a detection-bias with an intrinsically low probability of observing sources in that luminosity range}. The importance of a detection bias in the observation of intermediate-mass BHs has been thoroughly investigated. For example, \cite{Mezcua_2016_X} showed, by means of X-ray stacking, that a population of faintly accreting intermediate-mass BHs (with X-ray luminosity $L_{\rm x} \sim 10^{38}-10^{40} \, \mathrm{erg \, s^{-1}}$) should be present in dwarf galaxies; however, their detection is challenging due to their faintness and mild obscuration. Figure \ref{fig:LF} also suggests that the abundance of sources with $L_{\rm bol}< 10^{42} \, \mathrm{erg \, s^{-1}}$ should rise again at lower luminosities. We point out, though, that the detection of these abundant faint sources is challenging, because they enter the luminosity regime of stellar X-ray binaries \citep{Mezcua_2018_b}.

\section{Discussion and Conclusions} \label{sec:disc_concl}
We have employed a semi-analytical model to investigate the low-mass regime of the $\Mblack-\sigma$ relation and of the quasar luminosity function. Our model is based on two main assumptions: (i) low-mass and high-mass seed populations form at $z \sim 20$; (ii) the accretion process is bimodal, with BHs with $\Mblack \lesssim 10^5 \Msun$ accreting inefficiently. 

The $\Mblack-\sigma$ relation is somewhat tricky for low-mass galaxies, as $\sigma$ is defined as the velocity dispersion of stars inside bulges. An alternative to the $\Mblack-\sigma$ relation would be the $\Mblack-M_{\star}$ relation, where $M_{\star}$ is the stellar mass of the galaxy \citep{Reines_Volonteri_2015}. We note, however, that many dwarf galaxies do have bulges (e.g., NGC 4395, POX52, RGG 118, see \citealt{Baldassare_2015}) and that the definition of $\sigma$ can always be interpreted as the velocity dispersion of stars within some effective radius from the center of mass of the system.
In this Letter we chose to focus on the $\Mblack-\sigma$ because it seems to provide a tighter relation (e.g., \citealt{Shankar_2016}), indicating a more fundamental connection. As a test, we performed our simulations also in the $(\Mblack, M_{\star})$ parameter space, using the theoretical relation presented in \cite{Reines_Volonteri_2015}. We confirm, also in the $\Mblack-M_{\star}$ space, the presence of the same departure from the theoretical relation, occurring at $\Mblack \sim 10^5 \, \mathrm{\Msun}$.
Below we discuss the consequences of our results for BH seed models at high redshift.

\subsection{Observational predictions for BH seeding models}
Our main prediction is that central BHs and their host galaxies depart from the $\Mblack-\sigma$ relation for masses $\Mblack \lesssim 10^5 \, \mathrm{\Msun}$, becoming under-massive with respect to the extrapolation of the $\Mblack-\sigma$ to lower masses.
The $\Mblack-\sigma$ relation reflects, for $\Mblack \gtrsim 10^5 \Msun$, the connection between galaxy evolution and BH growth. The link is driven by the outflows generated by the BH energy and momentum output. Smaller BHs grow inefficiently and are unable to generate strong outflows triggering the growth-regulation process. For this reason, BHs with $\Mblack \lesssim 10^5 \Msun$ fail to reach the mass dictated by their velocity dispersion and become under-massive.
Previous observations (e.g., \citealt{Martin-Navarro_2018,MN_Mezcua_2018}) and simulations (e.g., \citealt{Alcazar_2017,Habouzit_2017}) already suggested that feedback is driven by BH activity for $\Mblack \gtrsim 10^5 \Msun$ and by supernova-driven winds for $\Mblack \lesssim 10^5 \Msun$.

Some papers (e.g., \citealt{Greene_2006, Mezcua_2017_review, MN_Mezcua_2018}) suggest an alternative scenario for the low-mass regime of the $\Mblack-\sigma$ relation, predicting a flattening at masses $\sim 10^5-10^6 \Msun$. \cite{MN_Mezcua_2018} explained this putative flattening with a weaker coupling between baryonic cooling and BH feedback, disconnecting the BH from the evolution of the host spheroid. Alternatively, the flattening could be explained with the prevalence of a high-mass formation channel for early seeds \citep{Volonteri_2010}. These high-mass seeds would fail to grow and just accumulate around their original mass, $\sim 10^5-10^6 \Msun$. In this Letter, we envisage that a flattening toward $\sim 10^5-10^6 \Msun$ would be due to an \textit{observational bias}. Observing BHs with $\Mblack \lesssim 10^5 \Msun$ is currently challenging, and the predicted paucity of BHs shining at $L_{\rm bol} \sim 10^{42} \, \mathrm{erg \, s^{-1}}$ (Sec. \ref{subsec:LF}) could add to this effect.
Instead of a flattening, our model clearly predicts a \textit{downward} departure from the $\Mblack-\sigma$ relation. 
\textit{A detection for $\Mblack \lesssim 10^5 \Msun$ of a relation of the form $\Mblack \sim \sigma^{\alpha}$ with $\alpha \gtrsim 7$ would be an important indicator of the existence of a bimodal population of BH seeds.}

Pushing the detection limit to $\Mblack \lesssim 10^4 \Msun$ \citep{Baldassare_2017,Chilingarian_2018} will enable to determine the relevance of the BH - galaxy connection for lower-mass galaxies, and whether or not a departure from the $\Mblack-\sigma$ relation occurs. A major role in this observational challenge will be played by future observatories, both in the electromagnetic (e.g., Lynx; see \citealt{Ben-Ami_2018}) and in the gravitational (e.g., LISA; see \citealt{LISA_2017}) realms.
This effort will not only shed light on the interconnection between BH and its host galaxy, but will ultimately provide important constraints on the seed formation mechanisms active in the high-redshift Universe. The observation in the local Universe of intermediate-mass BHs, as well as other proposed techniques (e.g., deriving the local super-massive BH occupation fraction, \citealt{Miller_2015}) will help us to understand processes occurred early in the history of the Universe.

\vspace{0.3cm}
F.P. acknowledges support from the NASA Chandra award No. AR8-19021A, and enlightening discussions with Vivienne Baldassare and Elena Gallo. 
M.M. acknowledges support from the Spanish Juan de la Cierva program (IJCI-2015-23944).
I.M.N. acknowledges support from the EU Marie Curie Global Fellowships.
This work was supported in part by the Black Hole Initiative at Harvard University, which is funded by a JTF grant.




%\begin{thebibliography}{}
%\end{thebibliography}

\bibliographystyle{mnras}
\bibliography{ms}
\label{lastpage}
\end{document}